\documentstyle[fancyheadings]{memsait}

\def \SAIT #1 #2 {{\em Mem.\ Soc.\ Astron.\ It.\/} {\bf #1}, #2}
\def \MESS #1 #2 {{\em The Messenger\/} {\bf #1}, #2}
\def \ASTRNACH #1 #2 {{\em Astron. Nach.\/} {\bf #1}, #2}
\def \AAP #1 #2 {{\em Astron. Astrophys.\/} {\bf #1}, #2}
\def \AAL #1 #2 {{\em Astron. Astrophys. Lett.\/} {\bf #1}, L#2}
\def \AAR #1 #2 {{\em Astron. Astrophys. Rev.\/} {\bf #1}, #2}
\def \AAS #1 #2 {{\em Astron. Astrophys. Suppl. Ser.\/} {\bf #1}, #2}
\def \AJ #1 #2 {{\em Astron. J.\/} {\bf #1}, #2}
\def \ANNREV #1 #2 {{\em Ann. Rev. Astron. Astrophys.\/} {\bf #1}, #2}
\def \APJ #1 #2 {{\em Astrophys. J.\/} {\bf #1}, #2}
\def \APJL #1 #2 {{\em Astrophys. J. Lett.\/} {\bf #1}, L#2}
\def \APJS #1 #2 {{\em Astrophys. J. Suppl.\/} {\bf #1}, #2}
\def \APSS #1 #2 {{\em Astrophys. Space Sci.\/} {\bf #1}, #2}
\def \ASR #1 #2 {{\em Adv. Space Res.\/} {\bf #1}, #2}
\def \BAIC #1 #2 {{\em Bull. Astron. Inst. Czechosl.\/} {\bf #1}, #2}
\def \JSQRT #1 #2 {{\em J. Quant. Spectrosc. Radiat. Transfer\/} {\bf #1}, #2}
\def \MN #1 #2 {{\em Mon. Not. R. S. Astr. Soc.\/} {\bf #1}, #2}
\def \MEM #1 #2 {{\em Mem. R. Astr. Soc.\/} {\bf #1}, #2}
\def \PLR #1 #2 {{\em Phys. Lett. Rev.\/} {\bf #1}, #2}
\def \PASJ #1 #2 {{\em Publ. Astron. Soc. Japan\/} {\bf #1}, #2}
\def \PASP #1 #2 {{\em Publ. Astr. Soc. Pacific\/} {\bf #1}, #2}
\def \NAT #1 #2 {{\em Nature\/} {\bf #1}, #2}

\input epsf.sty
\begin{opening}
\title{WHICH  THERMAL  PHYSICS FOR GRAVITATIONALLY UNSTABLE MEDIA?}

\author{Daniel Pfenniger}
\institute{Geneva Observatory \& University of Geneva, Geneva, Switzerland}

\date{} 
\end{opening}

\begin{document}

\oddpagefooter{\sf Mem. S.A.It., Vol. ??, ??}{}{\thepage}
\evenpagefooter{\thepage}{}{\sf Mem. S.A.It., Vol. ??, ??}
\bigskip

\begin{abstract}
We remind that the assumptions almost universally adopted among
astronomers concerning the physics to use to describe rarefied cosmic
gases remain often without justifications, mainly because the long
range of gravitation invalidates the use of classical thermal physics.
In turn, without sufficiently good local thermal equilibrium,
macroscopic quantities, such as temperature and pressure, are {\it not
defined\/} and the fundamental assumption that locally the medium is
smoothed by ``molecular chaos'' to justify the use of differential
equations is not granted.  The highly inhomogeneous fractal state of
the interstellar gas is probably a plain symptom of the large
discrepancy between the available theoretical tools, predicting local
homogeneity after a few sound crossing times, and reality.  Such
fundamental problems begin to occur in optically thin media such as
stellar atmospheres, but become exacerbated in the interstellar
medium, in cooling flows, and in the post-recombination gas,
particularly when gravitation becomes energetically dominant, i.e.,
when the medium is Jeans unstable.
\end{abstract}

\section{Introduction}
The purpose of this paper is to remind that the over century old
assumptions of classical thermodynamics and statistical physics have
never been properly adapted to gravitating systems.  This problem is
central to astrophysics, because concepts such as temperature or
pressure are constantly used throughout observational and theoretical
astronomy, with often a poor perception of their limits.

Chemists, physicists or engineers use thermodynamics (i.e.,
thermostatics) and statistical mechanics everyday with often great
success.  They must be aware of the limits of these theoretical tools,
when dealing with out-of-equilibrium systems such as convective fluid
layers or biological systems.  These systems often require new tools,
to adapt from case to case.  Indeed, the variety of possible systems
out of thermal equilibrium is incomparably larger than the variety of
systems in thermal equilibrium.  Any system displaying long range
correlations should be regarded as improper to be described with
thermodynamics on the global scale.  Much efforts have been and are
currently made in statistical mechanics to describe and understand
``phase transitions'', the meaning of which is often merely that such
systems are in non-thermal states displaying long range correlations.

On the other hand astrophysicists were accustomed long ago (e.g.,
Kelvin in the late 19'th century) to apply thermal physics with
success in planetary and stellar conditions.  The assumption of local
thermodynamic equilibrium (LTE) may be very good inside solar type
stars, thus the thermal physics applies well and leads to successful
progresses.  However, workers in stellar atmospheres realized
early that when the photon mean free path is larger than the density
scale-length, physics becomes non-local and non-LTE corrections must
be applied.

By lack of better tools, and by natural extension, the known thermal
physics was applied by famous scientists such as Jeans, Eddington,
Str\"omgren and Spitzer to interstellar gas.  Often such pioneer
attempts were considered by their authors as exploratory work, where
LTE is assumed as a very simplified working hypothesis.  However,
since better theoretical tools continued to be lacking, it became
admitted among astronomers to use these assumptions or variants of
them (detailed balance), with decreasing awareness of their limits.
Today these strong hypotheses are very often adopted without any
cautionary remarks.  A rereading of classical textbooks (e.g., Landau
\& Lifchitz 1966, ``Statistical Physics'', in particular Chap.\ I) may
be refreshing on this point.  Much of the warnings reminded here are
already stated there.

\section{Thermostatics and Entropy Fluxes in Astrophysics}

Basically, what is sometimes called the ``fourth principle of
thermodynamics'' (Landsberg 1984), i.e., that macroscopic quantities
(such as energy) are either intensive or extensive (additive),
proportional to the $0^{\rm th}$ or $1^{\rm st}$ power of the volume
$V$, fails in systems with long range interactions.  Thus the domains
of astrophysics where the gravitational instability develops (i.e.,
when gravitational energy exceeds pressure) are particularly
concerned, because the gravitational energy grows as $G \rho_0^2
V^{5/3}$ in uniform media.  In such situations long range correlations
and matter motions induced by gravity propagate with the dynamical
time-scale $\sim 1/\sqrt{G\rho_0}$, comparable to the sound crossing
time: the medium, not even in mechanical equilibrium, is then unable
to thermalize, to relax faster than dynamics. There is then no
justification to continue to apply thermostatics.

Perhaps not coincidental, such states are precisely related to the
least understood situations in astrophysics: star formation, galaxy
formation, and first structure formation after recombination.  All
involve some kind of gravitational collapse developing singularities,
and various back reactions from the small scale physics stopping or
slowing down the growth of singularities.

According to these remarks, the structures occurring in astrophysics
could be classified in three categories.
\smallskip
\begin{enumerate}
\item The ones dominated by microscopic physics, such as small scale
objects from atoms, molecules, grains up to planetary or stellar
atmospheres which have little self-gravity.  In such systems
thermodynamics applies often well because the dominant interactions
are short range.
\item The systems in which thermodynamics competes with gravity, such
as stars, the cold interstellar medium, or the gas after
radiation-matter decoupling.  Here thermodynamics is sometimes
applicable when LTE is a good approximation (when the medium is
optically thick), but often not.
\item The systems dominated by gravity such as the planetary systems,
stellar clusters, well formed galaxies, and large scale structures.
Here thermodynamics is irrelevant at a global scale since dynamical
relaxation is much longer than the system age.
\end{enumerate}

\medskip
The general growth of structures and correlations in the Universe is
opposite to an approach to a thermal equilibrium.  It requires a
growth of phase space volume faster than the entropy production by
microscopic phenomena.  For a definition of entropy and limits, see
the Appendix.  In other words, the general growth of structures in
astrophysics requires states sufficiently out of thermal equilibrium,
which is actually what makes fields like astrophysics or biology
interesting.  The steady production in time of ever increasingly
complex structures, from large scale cosmological structures down to
galaxies, stars and biological systems, follows from the negentropy
(minus the entropy) generated by the universal expansion and cascading
down the scales.  In other words, as for biology, out of equilibrium
systems characterize astrophysics.

If we take a close example of structural growth, complex biological
structures are not produced by the Sun {\it energy\/} input, as often
stated, but by the entropy flow: fortunately every erg received by the
Earth from the Sun is re-radiated in the infrared, otherwise the Earth
temperature would become rapidly uncomfortable!  But since each UV
photon is degraded into several IR photons, which increases the global
entropy by increasing the number of degrees of freedom, works can be
produced.  In this non-equilibrium process living systems succeed to
capture negentropy, allowing them to complexify.

\section{Gravitationally Unstable Media}

An infinite uniform medium raises fundamental difficulties that were
apparent immediately after the discovery of the law of gravitation, as
discussed in a famous letter exchange in 1692--1693 between Richard
Bentley and Isaac Newton (Newton 1779).  By symmetry an infinite
uniform medium cancels exactly the attraction force at any point, but
this is in fact a subtraction of infinities, the medium on the left of
a point attracts it with an infinite force, as well as the medium on
the right.  In such situations the slightest perturbation to the
medium at large distances may implies large residual forces.
Therefore Bentley and Newton agreed that a God was required to
constantly ``stabilize'' the Universe, seen then as necessarily
eternal, infinite and uniform.

Such an intuitive notion of (in)stability was refined by Jeans (1902)
with the mathematical description of the concept of stability (much
developed earlier by Poincar\'e): Is a slight perturbation growing in
time or not?  If yes, how fast?  This led to the condition of linear
gravitational instability in uniform media with gas pressure, giving a
scale, the Jeans length or mass, above which the medium is
gravitationally unstable.  Real physical systems are not infinite,
therefore the Jeans criterion allows to distinguish between a box of
gas in the laboratory, or in space, the size of which is practically
infinite for thermodynamic applications, but perhaps not for the Jeans
length.  If the box size is much larger than the Jeans length the
Bentley-Newton paradox applies, which in modern terms means that the
medium is unstable, and cannot remain as such.

The linear stability analysis is rapidly unable to describe the
non-linear phase of the gravitational instability due to the strongly
chaotic nature of the problem. Some further description of the
non-linear phase is necessary.  The still most widespread mental model
of the non-linear phase of a gravitational instability, alive since
several decades, is the rapid and synchronous ``crystallization'' of
the unstable uniform medium into a myriad of blobs of similar sizes
given by the initial Jeans length. Subsequently the blobs are imagined
to collapse synchronously nearly as if they were isolated spherical
blobs.

In fact numerical simulations of the non-linear phases in cosmological
and galactic contexts show a very different typical behaviour.  The
unstable medium develops pancakes, filaments and clusters over
time-scales much longer than the initial spherical collapse
time. Filamentary structuration proceeds, and clumps are rarely
isolated for long. Interactions, collisions, merging, but also
disruptions and evaporation of clumps are frequent. Contrary to
matter, voids do tend toward spherical shapes.  The whole structure
remains out of dynamical equilibrium as long as the large-scale
ordered motion (the Hubble flow, the galaxy differential rotation)
feeds the smaller scales.  The smallest scales may dissipate energy
and consume negentropy (i.e., they produce entropy).  Eventually some
scale-invariant order, scaling laws may emerge from such out
of equilibrium non-linear processes.

On may note that pancakes, filaments and clusters do occur in the
computer without including any radiative cooling.  This is important
because it shows that gravity alone is able to develop dense states
without requiring any atomic cooling phenomena.  This is related to
the ``negative specific heat'' of gravitational systems.  In general,
systems with negative specific heat are thermally unstable and tend to
develop spontaneously large fluctuations in temperature or density.

A nice illustration of the development and persistence of large scale
fractal-like structures is provided by the shearing sheet experiments
aimed at representing a small patch of a differentially rotating
self-gravitating disk (Wisdom \& Tremaine 1988; Toomre \& Kalnajs
1991; Salo 1992, 1995).  The anti-thermodynamic persistence of long
range correlations must be understood by the continuous flow of order
from large-scale {\it time-dependent\/} and correlated motion down to
small-scale dissipative and chaotic dynamics. In this case
gravitational instability does not lead to ``collapse" as long as the
system, driven by the large scale time-dependence of the boundary
conditions, absorbs at small scale the negentropy by chaos and
dissipative terms.

This aspect is very important to understand because usually the simple
minded picture of gravitational instability considers collapsing blobs
in isolation developing almost everywhere as the main outcome of
gravitational instability (thus, it is often believed that stars
should form in large amount in the early Universe).  Here we argue
that a better understanding of the non-linear regime of gravitational
instability leads to the possibility that point-like collapses are not
necessarily widespread, but that long range fractal order can persist
as long as the largest scale boundary conditions are time-dependent.
The main examples supporting this view are:
\smallskip
\begin{enumerate}
\item 
The fractal-like structure of galaxy and galaxy cluster distribution
(e.g., Coleman \& Pietronero 1992) over some range in scale, fed in
our view by the universal expansion.
\item 
The fractal structure of the interstellar medium (Scalo 1985) even
when not perturbed by star formation, is fed by galaxy differential
rotation (Fleck 1981).  This large-scale ordered motion cascades down
the cold interstellar clouds, which tend to follow scaling laws
(e.g., the size--velocity-dispersion relation, Larson 1981).
\end{enumerate}
\medskip

In these situations, scaling laws may persist as long as the larger
scale is driving the gravitational instability.  The boundary
conditions being time-dependent, a steady flow of entropy can
self-organize the system with a statistical quasi-steady state with
scaling laws.  The Kolmogorov picture of turbulence in incompressible
fluids is similar, except that cosmic gases are necessarily highly
compressible (since motion is trans- or supersonic), and the force
responsible to degrade the large-scale order by a cascade of
``eddies'' is gravity instead of molecular forces.

\begin{figure}
\parbox[]{7.5 cm}{\epsfxsize=7.5cm \epsfbox{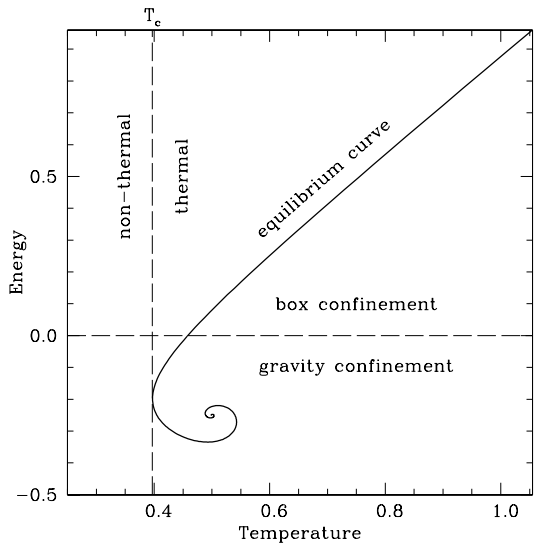}}%
\hfill\parbox[]{4.1cm}{\caption[t]{Equilibrium curve of the non-singular
isothermal self-gravitating gas sphere in an energy--temperature
diagram (dimensionless units, see Binney \& Tremaine 1987,
p. 503). The upper solid part corresponds to stable equilibrium, while
the lower, spiral part is mostly unstable.  The domain $T< T_{\rm c} =
0.397$ possesses no thermal and dynamical equilibrium, therefore the
medium is thermally and gravitationally unstable.\\ }}
\end{figure}

\section{The Isothermal Gravitating Gas Sphere and Beyond}

The limits of classical thermodynamics are particularly clear when
using a simple model of gravitating gas as close as possible to the
perfect gas box at a temperature fixed by a heat bath, used
extensively in thermodynamics.  To simplify the treatment a spherical
box is adopted (Lynden-Bell \& Wood 1968).  Such a configuration does
allow an equilibrium when the temperature is sufficiently high (Fig.\
1).  The high temperature limit tends toward the usual perfect gas
without self-gravity, where the gas confinement is due to the solid
walls of the box.  When the temperature drops, the total energy of the
gas decreases, and reaches negative values: the gas is then confined
mostly by self-gravity, and secondly by the box.  This is the regime
analogous to stars.  Yet, in this simple model nothing forbids to drop
the box temperature even more.  But then a last equilibrium model is
reached at $ T_{\rm c}\approx 0.397$, which corresponds to the Jeans
critical temperature for infinite uniform gas.  Below this critical
temperature {\it no thermal or dynamical equilibrium exists}.

Real cosmic gases are not perfect.  At sub-critical temperature one
expects wild fluctuations and the growth of singularities,
``revealing'' at some point new physical ingredients. By necessity an
eventual asymptotic statistical state, if it exists, must then depend
directly on the properties of the new ingredients revealed by the
growth of singularities, and on gravity.  The triggering of nuclear
reactions in stars is an example of possible new physical ingredients,
leading to long-term out-of-equilibrium situations, but other outcomes
could be considered, such as planets or black-holes.

Among the most necessary modifications of the isolated isothermal
perfect gas to represent cosmic gas, one may include energy
dissipation and time-dependent boundary conditions, since interstellar
clouds are far from being energetically isolated.  Indeed cosmic gas
cools rapidly, is often optically thin, so energy conservation is bad
in practice.  And molecular clouds (whose long term persistence is
still mysterious) are known to collide with substantial strength with
a time-scale similar to their internal crossing time.  So eventually
models including these aspects might explain better the statistical
properties of cosmic clouds.  One can show that if a gravitational
system finds a hierarchical order with fractal dimension $D$, clumps
tend to increase/decrease their kinetic ``temperature'' with size for
$D\,{^{>}_{<}}\,1$, which might thus be a criterion for star
formation.  Further, a steady energy flow across the scales occurs for
$D=5/3\approx 1.67$ (Pfenniger \& Combes 1994).

Numerical experiments with dissipative $N$-body systems are underway
and will be reported elsewhere.  It suffices to say here that, as the
shearing sheet experiments, 3D $N$-body systems can also find some
kind of statistical state with fractal dimension $D \sim 1-2$ when
ordered energy is injected at large-scale (representing cloud
collisions), and dissipated at small scale (by particle inelastic
collisions).  This is not fundamentally different from the situation
where a box in an external gravity field and half-filled with sand, is
periodically shaken.  The sand may quickly adopt spontaneously a well
defined vertical distribution unrelated to thermostatics, but with
balancing the macroscopic energy and entropy fluxes from the box scale
down to the grain scale.

\section{Conclusions}

The frequent assumption that cosmic gas is near a thermal equilibrium
is particularly questionable in the case of gravitationally unstable
media in which the basic requirements for applying thermodynamics are
not met (e.g., additivity of energy).  Indeed, as long as the medium
is gravitationally unstable, the speed of matter disturbances is 
comparable to the sound speed, and thermalization cannot be established.

Instead of classical thermodynamics one should tend to use what has
been learned in non-equilibrium statistical mechanics.  Scaling laws
are ubiquitous in open systems in which entropy flows.  We suggest
that in many situations gravitationally unstable media can reach a
dynamical equilibrium which transforms large scale ordered motion into
small scale ``turbulence''.  A fractal state can be maintained by
interaction with higher scales as long as the largest scale is not in
equilibrium, and energy and negentropy are dissipated at small scale.

The main astrophysical situations where one might expect the formation
of fractal states are: 1. At the formation of the first structures in
the Hubble flow just after recombination.  2. In cooling flows in
galaxies and galaxy clusters.  3. In spiral galaxies as long as the
amount of gas is sufficient to allow efficient cooling.  4.
Similarly, in other smaller self-gravitating gaseous disks.

The conventional isolation hypothesis in scenarios of star and galaxy
formation is probably inappropriate.  Much can be changed in the way
of understanding these processes if the interactions between upper and
lower scales are taken into account.  In particular, we see no ground
to necessarily expect an intense phase of star formation just after
recombination just because the medium becomes gravitationally
unstable.

\section*{Appendix: Entropy in Gravitating Systems}

A huge literature exists discussing ``entropy'' in widely different
contexts.  However much confusion occurs in astrophysics because some
forms of entropy are used, in particular Boltzmann's form, without
consideration whether the original assumptions (e.g., additivity of
energy) are applicable.

So here is a limited attempt to clarify the situation, largely based
on the clear exposition in Landau \& Lifchitz (1966), with some
additional remarks, inspired by David Ruelle (oral communication),
including the modern notion of chaos.

Entropy has a clear interpretation for {\it bounded\/} dynamical systems,
\begin{equation}
\dot{\vec z} = \vec F(\vec z) \  ,
\end{equation}
where $z$ is the phase space vector, and $\vec F$ the right-hand side
of the ``equations of motion".  Hamiltonian systems are a particular
case of dynamical system.  Entropy is defined as,
\begin{equation}
S = k \log{\Omega \over \Omega\rlap{$_0$}}\ ,
\end{equation}
where $\Omega$ is the ``a priori available phase space volume'' during the
time $\Delta t$, $\Omega_0$ is a constant volume of phase space (e.g.,
an incompressible elementary volume $(2\pi\hbar)^{N}$, where $N$ is
the number of degrees of freedom), and $k$ a constant used to relate
$S$ to the temperature unit.  Actually, entropy is best viewed as a
pure number, the logarithm of a phase-space volume ratio, so often
$k=1$ is chosen.

The meaning of ``a priori available phase space'' becomes clearer when
considering the chaos of most dynamical systems.  In strictly
deterministic systems, or when $\Delta t$ is much smaller than the
relaxation time, the potentially available phase space is restricted
to just one elementary cell of phase space, fixed by the initial
conditions, so entropy is zero.  But real physical systems cannot be
modeled as deterministic, not necessarily because of quantum effects
as often argued, but because they are often very chaotic, i.e.,
sensitive to perturbations.

To illustrate this point, we take as example the classical
deterministic description of molecular motion in a gas, perturbed by
the minute gravitational perturbation of a single electron at 1 Gpc
distance.  The additional acceleration may look negligible ($\Delta a
\approx 6\cdot 10^{-90}\,\rm cm \,s^{-2}$).  The collision-time of
atmospheric molecules is of the order of $\tau = 10^{-12}\,\rm s$,
during which the molecule paths are deviated by the electron
gravitational force by $3\cdot10^{-114}\,\rm cm$.  Each collision,
because of uncorrelated directions, doubles the previous {\it
squared\/} deviation, such that $|\Delta \vec x|^2 (n\tau) \approx 2^{n-1}
|\Delta \vec x|^2(\tau)$, so after only $10^{-9}\,\rm s$, $n=10^3$,
and $[|\Delta \vec x|^2 (1000\tau)]^{1/2} \approx 10^{37} \,
\rm cm$, very large indeed!  This is sufficiently huge to be obliged
to drop very quickly the assumption of strict isolation.  Thus for
such chaotic systems a probabilistic attitude must be taken.  The
``molecular chaos" invoked by Boltzmann is fully justified when taking
into account the chaotic nature, in the modern sense of chaos, of most
systems with many degrees of freedom.

Thus the exponential growth of perturbations compels to take into
account in realistic models of microscopic dynamics the uncontrolled
minute terms, for example in the form of a stochastic component
$\epsilon \, \vec \zeta(\vec z)$,
\begin{equation}
\dot{\vec z} = \vec F(\vec z) + \epsilon \, \vec \zeta(\vec z)\ ,
\end{equation}
which represents all the minute forces that by necessity cannot be
explicitly included in the system, but which leads to non-negligible
deviations of $\vec z$ over $\Delta t$.

In a dynamical system subject to perturbations by stochastic forces
the potentially accessible phase space is in fact rapidly the whole
chaotic region of the unperturbed system.  The notion of entropy
therefore is particularly useful in systems in which nearly all the
phase space is chaotic. The measure of entropy is then well
approximated by the whole phase space, and the ergodic hypothesis is a
good approximation.

In integrable or nearly integrable dynamical systems the available
phase-space due to stochastic perturbations may grow much slower in
regular than in chaotic regions.  If the time to have a sizable
probability to visit all parts of phase-space due to the stochastic
forces is much longer than the observation time $\Delta t$, the system
looks predictable and no statistical physics is required.  A good case
is the Solar System which in the present state is well described by
deterministic equations over less than a few Myr (Laskar 1994), but for
longer time-scales the most chaotic variables (associated with the
largest positive Liapunov exponents) must be considered in a
statistical sense.

The general definition of entropy in Equ.\ (2) can be applied to
non-additive bounded dynamical systems, while Boltzmann's entropy,
$
S_B = -k \sum_i  p_i \log p_i,
$
where $p_i$ is the probability of state $i$, requires a system with
{\em additive energy} (see Landau \& Lifchitz, 1966, \S 7).  So
Boltzmann entropy should not be used in gravitating systems.

\acknowledgements 
This work has been supported by the Swiss Science Foundation.

\end{document}